\documentclass[amsmath,amssymb,aps,reprint,groupedaddress]{revtex4-1}

\usepackage{graphicx}
\usepackage{dcolumn}
\usepackage{bm}
\usepackage{epstopdf}
\usepackage[flushleft]{threeparttable}
\AppendGraphicsExtensions{.tif}

\begin{document}


\title{Roles of Nitrogen Substitution and Surface Reconstruction on Stabilizing Non-passivated Single-layer Diamond}


\author{T. Pakornchote$^{a, b}$}
\author{A. Ektarawong$^{a, b}$}
\author{W. Busayaporn$^c$}
\author{U. Pinsook$^{a, b}$}
\author{T. Bovornratanaraks$^{a, b}$}
\email{thiti.b@chula.ac.th}
\affiliation{$^a$Extreme Conditions Physics Research Laboratory, Physics of Energy Materials Research Unit, Department of Physics, Faculty of Science, Chulalongkorn University, Bangkok, Thailand}
\affiliation{$^b$Thailand Center of Excellence in Physics, Commission on Higher Education, 328 Si Ayutthaya Road, Bangkok, 10400, Thailand}
\affiliation{$^c$Synchrotron Light Research Institute (Public Organization), Nakhon Ratchasima, 30000, Thailand}


\date{\today}

\begin{abstract}
The existence of single-layer diamond or diamane, which could adopt the properties of its bulk counterpart, has been verified by previous calculations and experiments. Even though, carbon atoms on the top and bottom surfaces need to form dangling bonds with atoms and/or molecules to stabilize their sp$^3$ hybridization. In this work, diamane is substituted by N atoms assisting the diamond-like structure to be stabilize without any passivation studied by \textit{ab-initio} calculation. One fourth of N substitution on diamane, vertically stacking as NCCC, is found to be stable by surface reconstruction forming Pandey $\pi$ chain structure and has an antiferromagnetic property. Half nitrogen substitution on diamane, vertically stacking as NCCN can be stable and prevails the diamane form. Its elastic constants yield high values, for example, its $C_{11}$ is as twice as diamond and its $C_{33}$ is only 21\% lower than diamond. The NCCN phase is a metastable phase at 10 GPa compared with other layered carbon nitride phases showing the possible pathway to be created.
\end{abstract}


\maketitle

\section{Introduction}
A single-layer diamond or diamane is prospective to be the thinnest hardness material. It can be created by tipping on bilayer graphene (BLG) to create the sp$^3$-bonding between interlayer carbon atoms enhancing the hardness of its substrate \cite{gao2018,barboza2011}. A layer of sp$^3$ carbon, however, cannot sustain its structure without tipping force \cite{gao2018}. The well-known problem is, for example, a surface reconstruction of bulk diamond. The surface of bulk diamond is either buckling, reconstructing the surficial layer or terminated by atoms and/or molecules in order to lower the surface energy \cite{davidson1994,pandey1982,scholze1996,lee1993,freedman1990,freedman1993}. For diamane, the carbon atoms at the top and bottom layers have to be passivated by atoms and/or molecules, \textit{i.e.} H, OH and F, in order to stabilize diamane making the passivated diamane metastable \cite{chernozatonskii2009,kvashnin2014,kvashnin2017,pakornchote2019,martins2017}. Diamane, whose dangling bonds on the surface are passivated by OH molecules, could be created at 5 GPa \cite{martins2017} same as diamond which typically becomes a stable phase at high pressure.

Martins \textit{et al.} \cite{martins2017} used multiwavelength Raman spectroscopy to show that G peak of BLG is discrepant in frequency between two wavelength lasers at 5 GPa. The results state that the dangling bonds of C atoms form sp$^3$ hybridization by passivating with OH, however, this transition occurs in partial area of the sample while the other areas was still sp$^2$ hybridization. The sp$^3$ carbon did not sustain when the sample was retrieved to ambient pressure. The passivated diamane can also be synthesized at ambient pressure by hot filament and chemical vapor deposition (CVD) techniques \cite{piazza2019,bakharev2020}. The stability is better if the passivation is on both side of diamane while one-side passivation is also possible for fewlayer graphene \cite{piazza2020}. The fluorinated diamane has been successfully synthesized by treating fluorine onto BLG for 12 hours. The time is sufficient for F atoms penetrating through BLG, so the fluorinated diamane can be formed by two-side passivation as shown by the image from scanning electron microscope \cite{bakharev2020}.

In the present work, N atom, which has one valence electron more than C atom, is considered to substitute in diamane instead of the passivation. Rather than hydrogenation or fluorination, N substitution in diamane is a 2-dimensional (2D) carbon nitride which could adopt the superhard property like in bulk carbon nitrides. Diamane is a diamond thinned down in [111] direction until has four C atoms in a unit cell by vertically stacking as CCCC leaving two lone electrons on the surface. The substitution of N atom on surficial C atom automatically create the lone electron pair and could stabilize the diamane-like structure. Four configurations of N substituted diamane (C$_{4-x}$N$_x$), derived from the unit cell of diamane with $x=$~1 and 2 by vertically stacking as NCCC, CNCC, NCCN and CNCN, are herein considered. Some of their intrinsic properties \textit{e.g.} elastic constants, electronic band structures and density of states (DOS), magnetism and phase transitions will be investigated and discussed.

\section{Computational method}
The density functional theory implemented in Vienna \textit{Ab initio} Simulation Package (VASP) was employed using Projector augmented wave (PAW) method for pseudopotential \cite{kresse1996-1,kresse1996-2,blochl1994-1,blochl1994-2}. Perdew-Burke-Ernzerhof (PBE) exchange-correlation functional was used for relaxing crystal structures, and Heyd-Scuseria-Ernzerhof (HSE06) screen exchange hybrid functional was used for calculating the energy gap by using the relaxed structures from PBE \cite{perdew1996,krukau2006}. In order to ensure the convergence of the energy, the cutoff energy was set as 600 eV, and the k-points were meshed using Monkhorst-Pack scheme \cite{monkhorst1976,pack1977}. The van der Waals correction using DFT-D3 method of Grimme was included for all calculation, presented in this work \cite{grimme2010}. To solve for phonon and vibrational scheme using finite displacement method implemented in PHONOPY package \cite{togo2008,togo2015}, VASP was used to calculate forces on each atomic in deviated crystal with 5$\times$5$\times$1 and 3$\times$3$\times$3 supercells for single-layer and bulk systems, respectively. The previous study shows that the magnetism of diamond surface is absent when using PBE, but present when using PBE0 and HSE06 \cite{pamuk2019}. Therefore, in this work, the collinear spin-polarized method together with HSE06 was employed to study the magnetic property of N substituted diamane.

\begin{figure}
\includegraphics[width=0.5\textwidth]{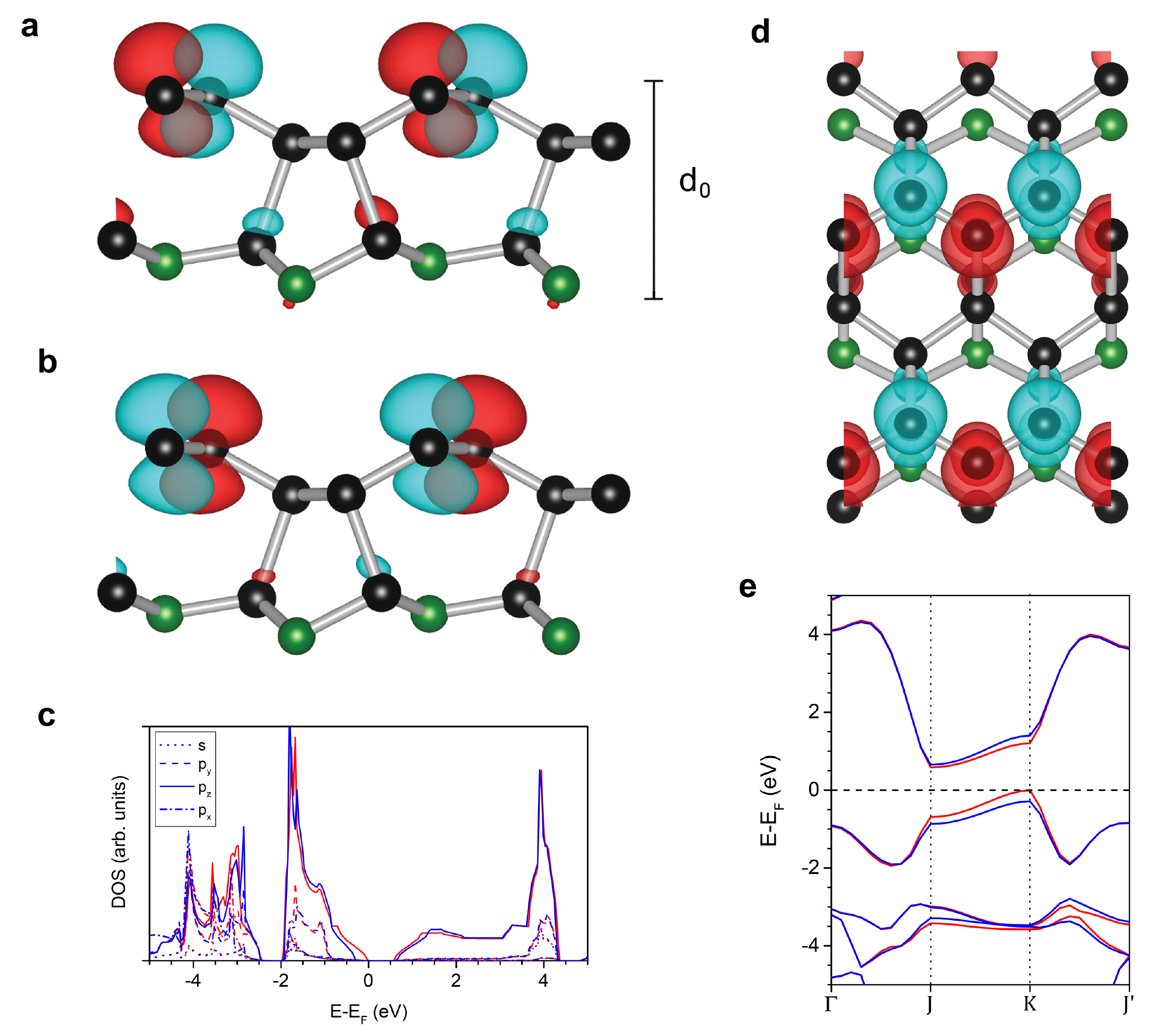}
\caption{\label{fig:pandey} The relaxed structure of $\pi$-C$_3$N in (a) [100] direction and (d) [001] direction where the blue and red envelopes are positive and negative magnetization densities, respectively, of valence band maximum at K-point and (b) of conduction band minimum at J-point. (c) Partial DOS of C atoms in $\pi$-C$_3$N and (e) band dispersion present the deviation between spin up (blue) and spin down (red). Black and green balls represent C and N atoms, respectively.}
\end{figure}

\section{Results and Discussion}

\subsection{Structures and Stabilities}
Certainly, diamane, which has four C atoms vertically stacking as CCCC, transforms to BLG after structural relaxation. Two of diamane substituted by one N atom (C$_3$N) vertically stacking as NCCC and CNCC also transform to a layer of graphene and CN after structural relaxation. Therefore, for NCCC, one side of N substitution forming one side electron lone pairs is not sufficient to energy of the system to be local minimum and stabilize the sp$3$ bonding between layers. A graphene side is then reconstructed to be the Pandey $\pi$ chain using a (2$\times$1)-rectangular cell (see Fig.~\ref{fig:pandey}) in order to lower surface energy from lone electrons on the surface \cite{pandey1982}. In contrast to diamane form, the reconstructed NCCC ($\pi$-C$_3$N) does not transform back to a layer of graphene and CN after structure relaxation. It is also dynamically stable verified by phonon calculated using PBE functional where the vibrational frequencies are listed in Supplemental Material. It is worth to note that the one-side Pandey $\pi$ chain forms of CCCC and CNCC are not stable by separating to two layers after structural relaxation.

The magnetic property of $\pi$-C$_3$N is investigated using collinear spin polarized calculation with PBE and HSE06 functionals resulting in absence and presence of magnetism in $\pi$-C$_3$N, respectively, in agreement with Ref.~\cite{pamuk2019}. The magnetic phase of $\pi$-C$_3$N has an energy 5.8 meV lower than its non-magnetic phase. Moreover, $\pi$-C$_3$N is an antiferromagnetic with zero total magnetization in contrast to the diamond surface which is a ferrimagnetic. The majority of magnetization is from top two C atoms obtaining 0.130$\mu_B$ and -0.128$\mu_B$. The valence band maximums of spin up and spin down largely split along the J-to-K path (see Fig.~\ref{fig:pandey}(e)). The energy gaps from spin up and spin down are indirect from a ground state at K-point to an excited state at J-point with 0.93 and 0.59 eV, respectively. The valance states on J-to-K path are from p$_z$-orbital depicted as a partial charge density adopting a shape of $\pi$-bonding envelope (see Fig.~\ref{fig:pandey}). The envelopes of spin up and spin down charge densities are at different and adjacent C atoms resulting to alternate direction of magnetization on the adjacent C atoms. The partial charge density of conduction band minimum at J-point shows a feature of $\pi$-antibonding where the charge density of spin up and spin down are around C atoms alternating to the valence band maximum at K-point.

\begin{figure}
\includegraphics[width=0.5\textwidth]{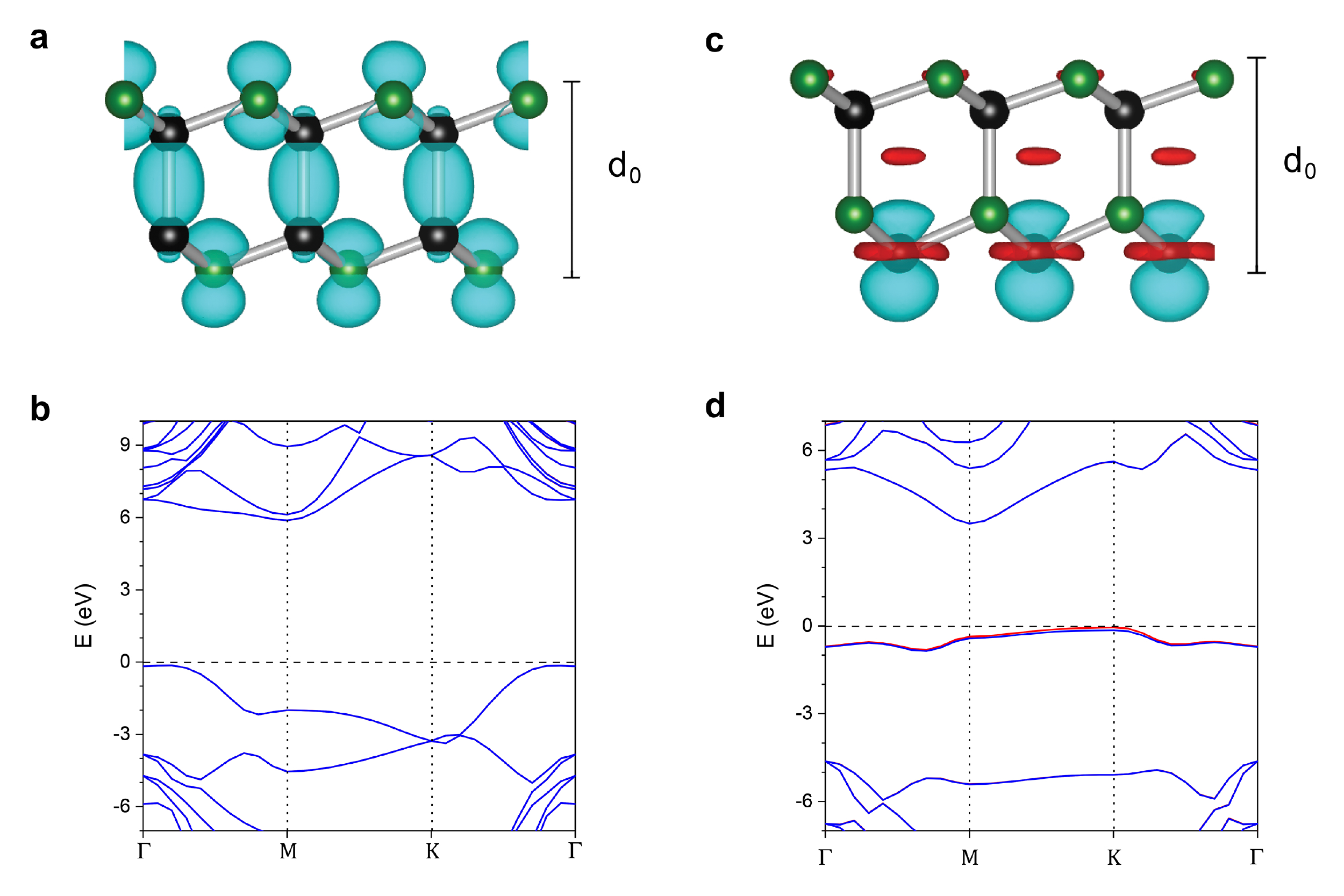}
\caption{\label{fig:cn} (a) The relaxed structure of NCCN is presented by projecting on [100] direction where blue envelopes are spin up charge density of valence band maximum at $\Gamma$-point. The relaxed structure of CNCN is presented by projecting on [100] direction where blue and red envelopes are positive and negative magnetization density, respectively, of valence band maximum at K-point. (b) and (d) are band structures of NCCN and CNCN, respectively. Black and green balls represent C and N atoms, respectively.}
\end{figure}

For C$_2$N$_2$, two configurations of N substituted diamane vertically stacking as NCCN and CNCN are studied (see Fig.~\ref{fig:cn}(a,c)). NCCN and CNCN are energetically stable after structural relaxation. However, only NCCN is dynamically stable. Moreover, collinear spin-polarized calculation with HSE06 functional is employed in order to determine the magnetism of NCCN and CNCN. In contrast to $\pi$-C$_3$N, the magnetism of NCCN is absence and CNCN has a tiny magnetization, 0.007$\mu_B$. For CNCN, the magnetism mainly arises from C atoms. The surficial C atoms and inner C atoms have magnetization about 0.008$\mu_B$ and -0.001$\mu_B$, respectively, as depicted in Fig.~\ref{fig:cn}(d).

For NCCN, the valence states are from p$_z$-orbital and yield the same charge density between spin up and spin down as a result in absence of magnetism. For CNCN, the valence states are from sp$_z$ hybridization. Its valence band maximum is nearly flat causing high density of states near the Fermi level and only this band that has a splitting between spin up and spin down (see Fig. S5 in Supplemental Material). The energy gaps are 6.1 and 4.4 eV for NCCN and CNCN, respectively. For CNCN, its bulk counterpart is dynamically stable, hence the single-layer phase could be dynamically stabilized if the carbon side sits on a substrate. This phase could be useful for some applications since it has very density of states at Fermi level with small magnetization.

\subsection{Elastic constants}
Some phases of carbon nitride were predicted to be superhard materials by which its cubic phase is even harder than diamond \cite{teter1996}. A bulk NCCN was shown that it has high elastic constant, $C_{11}$, however, it came with low $C_{33}$ \cite{bondarchuk2017}. In a perspective of 2D materials, BLG can resist the indented force better than graphite which is its bulk counterpart \cite{gao2018}. Since the volume of 2D materials, \textit{i.e.} single-layer graphene, is not well-defined, 2D elastic constant ($C^{\textrm{2D}}_{ij}$), which has N/m unit \cite{wei2009,ding2013}, evaluated with respect to pure C and N phases, and cen be expressed as
\begin{equation} \label{eq1}
C^{\textrm{2D}}_{ij} = \frac{1}{A_0}\Big(\frac{\partial^2E(\varepsilon_1,\ldots,\varepsilon_6)}{\partial\varepsilon_i\partial\varepsilon_j}\Big),
\end{equation}
where $A_0$ is the unstrained in-plane area of layered C$_x$N$_y$, and $\varepsilon_i$ is the applied strain up to $\pm2\%$ of the lattice parameters where $i=1, 2, …, 6$. In fact, the N substituted diamanes are not one atom thick, but it has semi-third dimension by which the sp$^3$ bonding in $z$-axis. Their elastic constants ($C_{ij}$) could be calculated by dividing with the effective height ($d_0$) which is a vertical distance between top and bottom atoms perpendicularly to the in-plane section,
\begin{equation} \label{eq2}
C_{ij} = \frac{1}{d_0A_0}\Big(\frac{\partial^2E(\varepsilon_1,\ldots,\varepsilon_6)}{\partial\varepsilon_i\partial\varepsilon_j}\Big),
\end{equation}
where $d_0=$~2.93, 2.59 and 2.67 {\AA} for $\pi$-C$_3$N, NCCN and CNCN, respectively. Because the N substituted diamanes have the effective thickness, so $C_{ij}$ in out-of-plane direction, \textit{e.g.} $C_{13}$, $C_{33}$, and $C_{44}$ are herein reported. The calculation by leaving $z$ direction unconstrained will give a bad result, then one may see the Supplemental Material for calculation procedure in VASP to obtain a reasonable result.

Table~\ref{tab:c2dij} presents the comparison of $C^{2D}_{ij}$ between several 2D materials. Obviously, single-layers of NCCN, CNCN, $\pi$-C$_3$N and hydrogenated diamane (H-diamane) have $C^{2D}_{ij}$ higher than single-layers of graphene and SiC \cite{pakornchote2019,ding2013,wei2009}. Therefore, the diamond-like structure yields $C^{2D}_{ij}$ higher than the flat structure. The reason is that $C^{2D}_{ij}$ is an intrinsic property for the flat structure since it lacks of third dimension, however, $C^{2D}_{ij}$ is an extrinsic property for the diamond-like structure. The energy used to stretch 2D materials is higher with thickness. Thus, those $C^{2D}_{ij}$ have to be divided by the thickness yielding $C_{ij}$.

Since the N substituted diamanes have a thickness from semi-third dimension, $C_{ij}$ were evaluated in order to compare with bulk materials. Table~\ref{tab:cij} lists the $C_{ij}$ of two phases of C$_2$N$_2$ in comparison with the bulk diamond. $C_{11}$, $C_{66}$ and $C_{12}$ of NCCN yield about two times higher than diamond and H-diamane while its $C_{33}$ is about 20\% lower than diamond and its $C_{44}$ is about half of that of diamond. CNCN yields similar result to NCCN. Therefore, NCCN and CNCN are tougher (softer) than diamond in in-plane (out-of-plane) direction and more vulnerable to shearing indicated by the comparison of $C_{44}$. We want to state that Eq.~\ref{eq2} could be just the upper limit of $C_{ij}$ where the lower limit is when $d_0$ is the height of the bulk form reported in Ref.~\cite{bondarchuk2017}. The Born stability criteria \cite{mouhat2014} was considered and found that both NCCN and CNCN are mechanically stable. For $\pi$-C$_3$N, its $C_{11}$ is as much as $C_{11}$ of NCCN, its $C_{22}$ is a bit smaller than $C_{22}$ of NCCN, and its $C_{33}$ is significantly lower than with $C_{33}$ of NCCN. It is worth to mark that $C_{22}$, a response in an armchair direction, is lower than $C_{11}$, a response in a zigzag direction. 

\begin{table}
\begin{threeparttable}
\caption{\label{tab:c2dij}The 2D elastic constants of single-layer NCCN, CNCN, $\pi$-C$_3$N, H-diamane, graphene and SiC in N/m unit.}
\begin{ruledtabular}
\begin{tabular}{cccccccc}
Phase&$C^{\textrm{2D}}_{11}$&$C^{\textrm{2D}}_{12}$&$C^{\textrm{2D}}_{13}$&$C^{\textrm{2D}}_{22}$&$C^{\textrm{2D}}_{33}$&$C^{\textrm{2D}}_{44}$&$C^{\textrm{2D}}_{66}$\\
\colrule
NCCN&568&66&51&&217&66&243\\
CNCN&526&61&38&&170&&220\\
$\pi$-C$_3$N&595&106&27&510&159&&244\\
H-diamane\tnote{a}&487&38&&&&&\\
Graphene\tnote{b}&358&60&&&&&\\
SiC\tnote{c}&179&57&&&&&\\
\end{tabular}
\end{ruledtabular}
\begin{tablenotes}
\item[a] Ref.~\cite{pakornchote2019}.
\item[b] Ref.~\cite{wei2009}.
\item[c] Ref.~\cite{ding2013}.
\end{tablenotes}
\end{threeparttable}
\end{table}

\begin{table}
\begin{threeparttable}
\caption{\label{tab:cij}The bulk elastic constants of single-layer phases which are NCCN, CNCN, $\pi$-C$_3$N and H-diamane, and their bulk counterparts in GPa unit.}
\begin{ruledtabular}
\begin{tabular}{ccccccccc}
&Phase&$C_{11}$&$C_{12}$&$C_{13}$&$C_{22}$&$C_{33}$&$C_{44}$&$C_{66}$\\
\colrule
Single-layer&NCCN&2191&253&198&&836&256&939\\
&CNCN&1968&227&142&&635&&825\\
&$\pi$-C$_3$N&2030&361&91&1739&541&&831\\
&H-diamane\tnote{a}&1026&81&&&&\\
\colrule
Bulk&NCCN\tnote{b}&965&107&-10&&59&10&430\\
&CNCN\tnote{b}&929&102&3&&73&37&414\\
&Diamond\tnote{c}&1051&127&127&&1061&559&559\\
\end{tabular}
\end{ruledtabular}
\begin{tablenotes}
\item[a] Ref.~\cite{pakornchote2019}.
\item[b] Ref.~\cite{bondarchuk2017}.
\item[c] Ref.~\cite{mcskimin1972}.
\end{tablenotes}
\end{threeparttable}
\end{table}

\subsection{Formation of enthalpy and Phase diagrams}
To pursue the possibility of further synthesis, the formation enthalpy ($\Delta H_{form}$) of C$_2$N$_2$ systems was evaluated by,
\begin{equation} \label{eq3}
\Delta H_{form} (\textrm{C}_x\textrm{N}_y) = H(\textrm{C}_x\textrm{N}_y) - \frac{xH(\textrm{C}) + yH(\textrm{N})}{x+y},
\end{equation}
where $x$ and $y$, respectively, are numbers of C and N atoms in the system, and the enthalpy terms on the right-hand side are the enthalpies of those systems per atom. Some most stable phases of carbon nitride are included in the phase diagram where they can be considered as two types; layered and non-layered morphologies. The layered phases of carbon nitride are g-C$_3$N$_4$, reconstructed g-C$_3$N$_4$ and polyheptazine (the structures are provided in Supplemental Material) that could be a starting material of N substituted diamanes. For the non-layered phases, the structures included in the phase diagram are $Pca2_1$-C$_3$N$_4$ and $P4_2/m$-CN, which are energetically favorable phases among their stoichiometries at pressures considered in the phase diagram \cite{pickard2016}.

\begin{figure}
\includegraphics[width=0.5\textwidth]{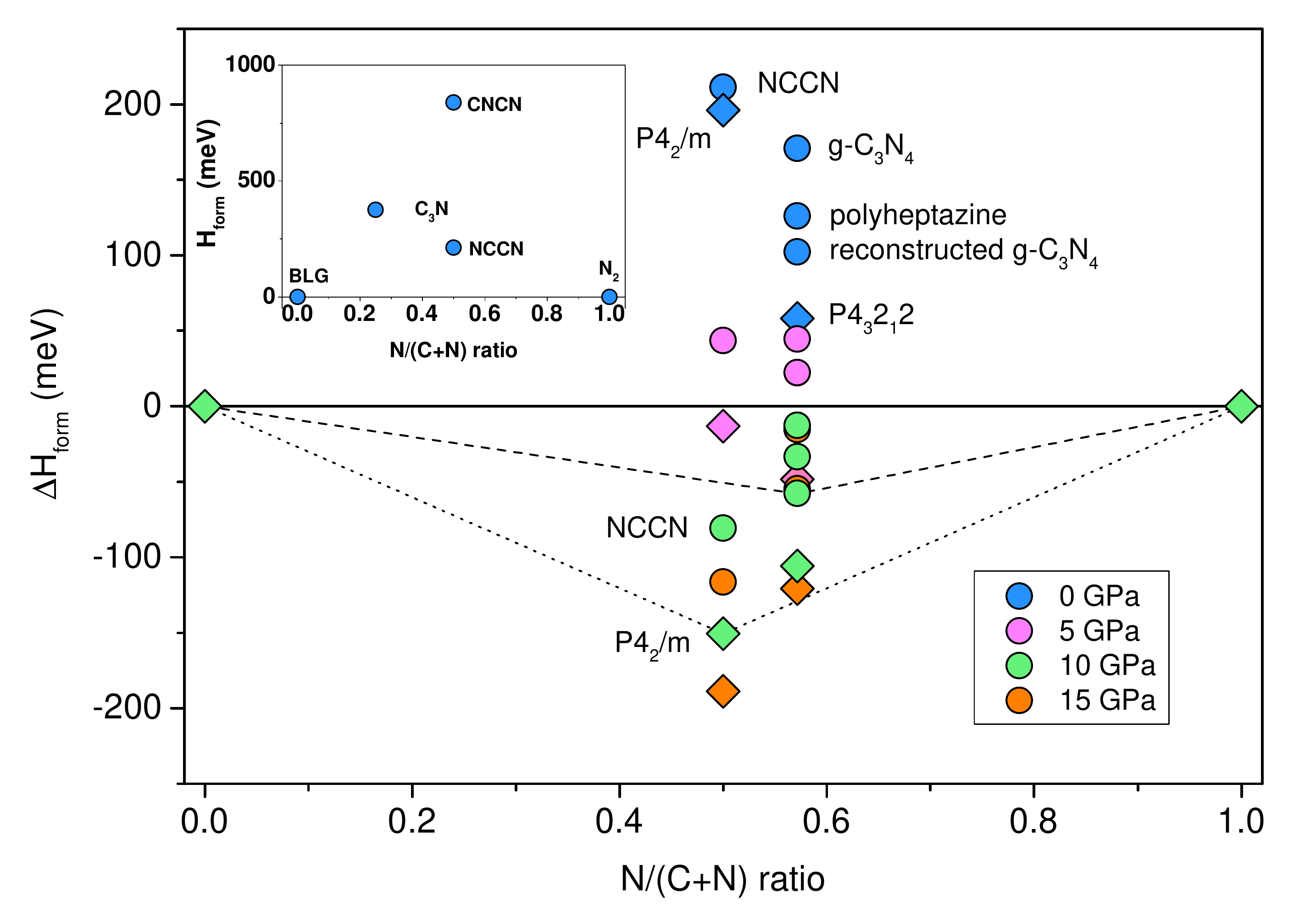}
\caption{\label{fig:convex}The formation enthalpy at various pressures. Circle and diamond symbols indicate to layered and non-layered phases, respectively. Dashed (Dotted) line is a convex hull connecting between diamond and reconstructed g-C$_3$N$_4$ ($P4_2/m$-CN) and $P4_12_12$-N$_2$.}
\end{figure}

The inset in Fig.~\ref{fig:convex} shows that $\Delta H_{form}$ at 0 GPa of NCCN, CNCN, $\pi$-C$_3$N are 211, 839, and 375 meV, respectively. Since their $\Delta H_{form}$ are even higher than other carbon nitrides (see Fig.~\ref{fig:convex}), these phases cannot be formed by using BLG or other carbon nitrides as precursors at ambient pressure. Likewise, diamond has enthalpy lower than graphite at 5 GPa, the study hence has been further investigated at high pressure. The referencing phases, changed from BLG and N$_2$ molecule, are diamond for pure C phase at 5-20 GPa, and for pure N phase, $P2_1/c$ phase at 5 GPa and $P4_12_12$ phase at 10-20 GPa \cite{pickard2009}. At ambient pressure, the N substituted diamanes are considered as single layers, however, at high pressure the compression in out-of-plane direction is inevitable and must be taken into account in the calculation. The $c$-axis is allowed to be relaxed in order to create isotropic pressure around materials trading off with that NCCN and others are no longer single-layer phase. 

Fig.~\ref{fig:convex} shows the phase diagram of carbon nitrides comparing with C phases and N phases at ambient and under high pressure. The CNCN and $\pi$-C$_3$N have $\Delta H_{form}$ higher than NCCN even at high pressure, so only NCCN will be herein discussed. As depicted in Fig.~\ref{fig:convex}, $\Delta H_{form}$ of NCCN decreases much from 211 meV at ambient pressure to 43.5 meV at 5 GPa and -83.3 meV at 10 GPa. The dashed lines are illustrated as a convex hull connecting with reconstructed g-C$_3$N$_4$ and showing that $\Delta H_{form}$ of NCCN is the lowest among layered phases at 10 GPa. Although, it is higher than non-layered phases, \textit{e.g.} $P4_2/m$ and $P4_32_12$ as illustrated a dotted convex hull. In the graphene case, it can be recovered from high pressure without buckling \cite{martins2017,smith2015,pakornchote2020}. Therefore, the phase transition to non-layered phases could be prevented if the choice of precursors is restricted to be 2D materials \textit{i.e.} bilayer g-C$_3$N$_4$. The phase diagram would have only layered phases left to consider.

\section{Conclusions}
In conclusion, we present the stabilization of the N subsitution in diamane. The substitution of one N atom in diamane, C$_3$N, is not sufficient to stabilize the diamane structure. The unsubstituted surface of NCCC must be reconstructed to be the Pandey $\pi$ chain in order to stabilize the bonding between layers. On the other hand, CNCC cannot be stabilized in any cases because the electrons may not form the lone electron pair on the N substituted side. Herein, $\pi$-C$_3$N and NCCN are shown that they can enhance the stabilization of diamond-like carbon film with non-passivation where NCCN is energetically favorable at 10 GPa comparing with the layered phases of carbon nitride. They also keep the high values of elastic constants as comparable as diamond which potentially present superhard property.

\begin{acknowledgments}
This research project was supported by the Second Century Fund (C2F), Chulalongkorn University. It is also partially supported by Super SCI-IV research grant, Faculty of Science and Ratchadaphiseksomphot Endowment Fund of Chulalongkorn University, Grant for Research. The authors would like to acknowledge the Computational Materials Physics (CMP) project, SLRI, Thailand for providing support on computational resource.
\end{acknowledgments}

\bibliography{bibfile}

\end{document}